\documentstyle[aps,twocolumn]{revtex}
\newcommand{\ds}{\displaystyle}
\newcommand{\ltwid}{\raise.3ex\hbox{$<$\kern-.75em\lower1ex\hbox{$\sim$}}}
\newcommand{\rtwid}{\raise.3ex\hbox{$>$\kern-.75em\lower1ex\hbox{$\sim$}}}
\renewcommand{\theequation}{\arabic{section}.\arabic{equation}}

\title{Superconducting Condensate Formation in Quasi-2D Systems
       with Arbitrary Carrier Density}

\author{Vadim~M.~Loktev$^{1}$,
Rachel~M.~Quick$^{2}$\thanks{Corresponding
author, e-mail: rcarter@scientia.up.ac.za}
and Sergei~G.~Sharapov$^{2}$\thanks{On leave of absence from
Bogolyubov Institute for Theoretical Physics, 252143 Kiev, Ukraine}
}

\address{$^{1}$Bogolyubov Institute for Theoretical Physics,
         252143 Kiev, Ukraine\\
         $^{2}$Department of Physics, University of Pretoria,
         0002 Pretoria, South Africa}

\date{October 29, 1998}

\begin{document}

\draft
\maketitle

\begin{abstract}
A phase diagram for a quasi-2D metal with variable carrier density
has been derived. The phases present are the normal phase, where the
order parameter is zero; the pseudogap phase where the absolute value
of the order parameter is non-zero but its phase is random, and a
superconducting phase with a crossover quasi-2D
Berezinskii-Kosterlitz-Thouless (BKT) region. The crossover region is
bounded by the quasi-2D BKT temperature and the temperature for the
onset of conventional long-range order (CLRO). The practical
observation of these regions however critically depends on the
carrier density.  At high densities both the pseudogap and BKT
regions vanish asymptotically i.e.  one obtains the standard BCS
picture. At intermediate densities the pseudogap phase is large but
the BKT region negligible. Finally at very low densities both the
pseudogap and BKT regions are sizeable.  An attempt is made to explain
the behaviour observed in underdoped (intermediate densities) and
optimally doped high-$T_{c}$ superconducting compounds above their
critical temperature. The transition to the pseudogap phase should
also be regarded as a crossover.
\end{abstract}

\nopagebreak

\section{Introduction}

It is widely accepted
\cite{Randeria.book,Emery,Loktev.review,Uemura,Randeria.1997,LSh.Review}
that the formation of the superconducting state in high-temperature
superconductors (HTSCs) is quite different from traditional
superconductors. The strongly anisotropic, almost two-dimensional,
structure of HTSCs compounds increases significantly the influence
of fluctuations and the different types of order appear to be
separated on  different energy scales. One type of order may correspond
to the formation of a {\em  pseudogap} (or lowered density of the
states at the Fermi level) above the critical temperature $T_{c}$
of the superconducting transition. The anomalous behaviour of HTSCs
\cite{Loktev.review,Pines.review} (including the behaviour of the spin
susceptibility, resistivity, specific heat and photo-emission spectra)
can then be interpreted in terms of the formation of this pseudogap
\cite{Randeria.1997,LSh.Review,Randeria.Nature}.

There are many approaches for studying these extremely complicated
systems. In most models, whether they are discrete (Hubbard) or
continuous, the systems are treated as two dimensional (2D). Even
the simplest 2D model with local nonretarded attraction between
carriers is still a very rich and complicated system. Its properties
beyond the mean field level are not well established, especially if
one interested in a wide range of carrier densities and coupling
strengths.

Most theoretical approaches (see, for example,
\cite{Micnas,Tchernyshyov} and the references therein) use the so
called self-consistent T-matrix approximation, where all Green
functions are "dressed". Although the self-consistency is an
important feature \cite{Tchernyshyov,Sofo,Serene,Levin,Nazarenko},
we stress that even this approximation cannot adequately describe
the formation of the Berezinskii-Kosterlitz-Thouless (BKT) phase
and it is well known that this is the only type of order possible
in 2D \cite{Coleman}.  The features of the superconducting BKT
transition were studied in \cite{Gusynin} although without
self-consistency.

It is important to note, however, that often real systems (in
particular HTSCs) are not strictly 2D systems and a pure 2D
scenario cannot be applied to them directly. The presence of a
third spatial direction already permits the formation of a
superconducting phase with long-range order (LRO). One cannot say
{\em apriori} that the formation of this phase destroys the 2D
physics abruptly. Nevertheless, using some estimations obtained
more than 20 years ago \cite{Kats,Efetov} for the quasi-1D metals,
in principle even for relatively modest anisotropies and carrier
densities the standard BCS behaviour should be recovered. The low
and intermediate density limit is not as clear and, given the
relevance of this limit to HTSCs with their small and variable
carrier density, it should be examined carefully.

The purpose of this paper is to build the complete phase diagram of a
quasi-2D metallic system with an arbitrary carrier density. We
have considered both the conventional long-range order (CLRO)
superconducting transition and the BKT transition. The accompanying
publications were devoted to the calculation of the temperature for
the BKT transition in the quasi-2D system \cite{QSh}, the temperature
of the CLRO transition in the extreme Bose limit \cite{GLSh.PhysicaC}
(see also review \cite{LSh.Review})).  To explain the origin of the
whole phase diagram and especially the part related to the pseudogap
region we recap here only the essential details from our previous works.
The calculation for the temperature of the CLRO transition in the
high density limit is new and we present here its full derivation
(see also a preprint of authors devoted to the Coleman-Weinberg
formalism in the theory of superconductivity \cite{QSh.preprint}).

Indeed a quasi-2D BKT transition is only physically meaningful
in the limit of weak three dimensionalization i.e. for a
practically two dimensional system. In this limit we argue that for
all carrier densities the BKT transition occurs before or at the CLRO
phase. Certainly in the cuprates, there was until very recently little or
no evidence for two superconducting phases and indeed it could be
argued from experiment that there was only one superconducting transition.
However in a recent paper \cite{Zverev} one indeed sees experimental
evidence in underdoped Bi-cuprates for the onset of planar superconductivity
before that of superconductivity perpendicular to the planes. This
experiment which gives a temperature difference of the order of 10K thus
supports the picture of the formation of a quasi-2D BKT region prior to the
formation of CLRO.

We also want to argue that the calculation of the critical temperatures for
the quasi-2D BKT phase and the CLRO phase is very different both physically
and mathematically.  The BKT crossover region has a 2D counterpart since
in two dimensions $T_{\rm BKT}$ is finite; the CLRO phase does not
since $T_c$ is zero \cite{Coleman}.  Moreover at very low (physically
unimportant) densities the transition temperatures are straightforward to
calculate and very different \cite{GLSh.PhysicaC}. The closeness of the
two temperatures in the physically important region (close to optimal
doping) is a non-trivial result since the calculations are different in
both the physical assumptions made and the mathematical details of the
approximations.  We find it remarkable, given the highly anisotropic
nature of the system, that the transition to conventional LRO takes
place at a temperature close to that of the BKT transition.

We speculate that the quasi-2D BKT region corresponds to the
crossover scale to three dimensional order. There may well
be no sharp transition at the critical temperature for BKT phase
formation.

More importantly, the superconducting region is separated from the
normal phase by the pseudogap phase with nonzero neutral order
parameter \cite{Gusynin,QSh}. Clearly the transition to this phase
should be a crossover but our approximations are insufficiently
accurate to give this behaviour (see also discussion in \cite{QSh}).
Furthermore this phase is two dimensional and large enough
to explain the observed anomalies when the density of the carriers is
relatively low (the underdoped case). Nonetheless the transition to
the superconducting phase is almost directly into a state with the
conventional three-dimensional long-range order at relevant doping
levels. Thus we have a crossover with increasing temperature from three
dimensional behaviour in the superconducting state to two dimensional
behaviour in the pseudogap phase similar to the picture recently proposed
by \cite{Vilk}.

The rest of the paper is organized as follows. In Sec.~II we
introduce the model and sketch the formalism applied. This
formalism is suitable for studying both the BKT and CLRO
transitions. The equations for $T_{\rm BKT}^{q}$, the temperature
for the BKT transition in the quasi-2D system, have been derived
previously \cite{QSh} and are simply sketched.  The CLRO
superconducting transition temperature for small and relatively
large carrier densities is derived in Sec.~III. In the high density
limit we go beyond the Nozi\`eres and Schmitt-Rink approach
\cite{Nozieres} which in the quasi-2D case, i.e. in the presence of
a third direction, simply gives the BCS critical temperature. By
employing the Coleman-Weinberg effective potential
\cite{EWeinberg1} we are able to calculate the Gaussian corrections
to the critical temperature and we obtain the correct limiting
behaviour $T_c \rightarrow 0$ as the system becomes increasingly
two-dimensional. The full phase diagram of the system is discussed
in Sec.~IV and Sec.~V presents our conclusions.

\section{Model and Formalism}
\setcounter{equation}{0}

Since the precise nature of the interplane tunneling in HTSCs is
not yet known \cite{Marel1} several different models exist. Here we
choose the simplest possible Hamiltonian density which is widely
used in the study of HTSCs \cite{Varlamov,GLSh.PhysicaC},
\begin{eqnarray}
H = & - & \psi_{\sigma}^{\dagger}(\mbox{\bf r})
\left[ \frac{\nabla_{\perp}^{2}}{2 m_{\perp}} +
\frac{1}{m_{z} d^{2}} \cos(i d \nabla_{z}) + \mu \right]
\psi_{\sigma}(\mbox{\bf r}) \nonumber
\\ & - & V
\psi_{\uparrow}^{\dagger}(\mbox{\bf r})
    \psi_{\downarrow}^{\dagger}(\mbox{\bf r}) \psi_{\downarrow}(\mbox{\bf r})
    \psi_{\uparrow}(\mbox{\bf r}),
\label{Hamiltonian}
\end{eqnarray}
where $\mbox{\bf r} \equiv \mbox{\bf r}_{\perp}, r_{z}$ (with
$\mbox{\bf r}_{\perp}$ being a 2D vector); $\psi_{\sigma}(\mbox{\bf r})$ is a
fermion field, $\sigma = \uparrow, \downarrow$ is the spin variable;
$m_{\perp}$ is the effective carrier mass in the planes (for example
CuO$_{2}$ planes); $m_{z}$ is an effective mass in the $z$-direction; $d$ is
the interlayer distance; $V$ is an effective local attraction constant; $\mu$
is a chemical potential which fixes the carrier density $n_{f}$; and we take
$\hbar = k_{B} = 1$ and $m_z >> m_\perp$.

In the weak coupling limit it is appropriate to replace the
attraction constant $V$ by the two-particle bound state energy in
vacuum \cite{Miyake,GGL},
\begin{equation}
\varepsilon_b = - 2W \exp \left( - \frac{4 \pi d}{m_\perp V} \right).
\label{bound.energy}
\end{equation}
Here $W$ is the width in the plane and the limit $V \rightarrow
0$, $W \rightarrow \infty$ is to be understood. This replacement
enables one to regularize the ultraviolet divergences in the
theory. One can then define the dimensionless system parameter
\begin{equation}
\tilde{\epsilon} = \frac{\epsilon_F}{|\varepsilon_b|}
\end{equation}
where  $\tilde{\epsilon} <<1$ corresponds to Bose or local pair
superconductivity and $\tilde{\epsilon} >>1$ to BCS
superconductivity. Since we have a quasi-2D system with a quadratic
dispersion law in the planes the Fermi energy $\epsilon_F$ is given
by
\begin{equation}
\epsilon_F = \frac{\pi n_F d}{m_\perp}.
\end{equation}
It was suggested in \cite{Carter} that optimally doped HTSCs have
$\tilde{\epsilon} \sim 3 \cdot 10^2-10^3$ while conventional
metallic superconductors have at least $\tilde{\epsilon}
\sim 10^3-10^4$.

The proposed Hamiltonian proves very convenient for the study of
fluctuation stabilization by weak 3D one-particle inter-plane
tunneling. The two particle (Josephson) tunneling has been omitted
in (\ref{Hamiltonian}) on the assumption that it is less important
than the one-particle coherent tunneling already included. In
certain situations however Josephson tunneling is more important.
In fact some authors consider the most important mechanism for
HTSCs to be the incoherent inter-plane hopping (through, for
instance, the impurity (localized) states or due to the assistance
of phonons). Nonetheless we omit Josephson tunnelling here. However
the single-particle dispersion relationship used here automatically
incorporates the layered structure of HTSCs which is a vital
extension to the commonly used 2D models.

It is significant that the large anisotropy in the conductivity
cannot be related to the anisotropy in the effective masses $m_{z}$
and $m_{\perp}$. In particular, HTSCs with rather large anisotropy
in the $z$-direction do not display conventional metallic behaviour
at low temperatures \cite{Watanabe}. However this semiconducting
behaviour is not directly related to the pseudogap phenomena
\cite{Watanabe} and the Hamiltonian (\ref{Hamiltonian}) may be used
to study the qualitative features of pseudogap opening.

The standard Hubbard-Stratonovich method was used to study the
Hamiltonian (\ref{Hamiltonian}). In this method the statistical sum
$Z(v,\mu,T)$ ($v$ is the volume of the system) is formally
rewritten as a functional integral over the auxiliary fields $\Phi
= V \psi_{\downarrow} \psi_{\uparrow}$ and $\Phi^{\ast} = V
\psi_{\uparrow}^{\dagger} \psi_{\downarrow}^{\dagger}$
\begin{equation}
Z(v,\mu,T) = \int {\cal D} \Phi {\cal D}
\Phi^{\ast} \exp[-\beta \Omega(v, \mu, T, \Phi(x), \Phi^{\ast}(x))],
                                \label{statsum}
\end{equation}
where
\begin{eqnarray}
&& \beta \Omega(v, \mu, T, \Phi(x), \Phi^{\ast}(x)) =
\nonumber                      \\
&& \frac{1}{V} \int_{0}^{\beta} d \tau \int d
\mbox{\bf r} |\Phi(x)|^{2} - \mbox{Tr Ln} G^{-1}[\Phi(x),\Phi^{\ast}(x)]
                                    \label{Effective.Action}
\end{eqnarray}
is the effective action and $x= \tau,\mbox{\bf r}$ denotes both the imaginary
time $\tau$ and the position $\mbox{\bf r}$ previously defined.  The action
(\ref{Effective.Action}) is expressed in terms of the Green function $G$
which has the following operator form
\begin{eqnarray}
&& G^{-1}[\Phi(x),\Phi^{\ast}(x)] =
\nonumber \\ && - \hat{I} \partial_{\tau} + \tau_{3}
\left[\frac{\nabla_{\perp}^{2}}{2 m_{\perp}} +
\frac{1}{m_{z} d^{2}} \cos(i d \nabla_{z}) + \mu \right]
\nonumber                  \\
&& + \tau_{+} \Phi(x) + \tau_{-} \Phi^{\ast}(x).
                     \label{Green.initial}
\end{eqnarray}
where $\tau_3, \tau_{\pm} = (\tau_1 \pm i \tau_2)/2$ are Pauli
matrices. Although the representation (\ref{statsum}),
(\ref{Effective.Action}) is exact, in practical calculations it
is necessary to restrict ourselves to some approximation.
The most convenient approximation for our purposes is the Coleman-Weinberg
\cite{EWeinberg1} (see also \cite{Miransky}) effective potential in the
one-loop approximation which we will use for studying LRO.  The exact
expression (\ref{statsum}) is replaced by
\begin{equation}
Z(v,\mu,T, |\Phi|^{2}) =
\exp[-\beta \Omega_{pot}(v, \mu, T, |\Phi|^{2})],
                          \label{statsum.Gaussian}
\end{equation}
where the effective thermodynamical
potential \begin{equation} \Omega_{pot}(v, \mu, T, |\Phi|^{2}) \simeq
\Omega_{pot}^{\mbox{\tiny MF}}(v, \mu, T, |\Phi|^{2}) +
\Omega^{(1)}(v, \mu, T, |\Phi|^{2})
                                    \label{Effective.Potential}
\end{equation}
is expressed through the mean-field "tree-potential"
\begin{equation}
\Omega_{pot}^{\mbox{\tiny MF}}(v, \mu, T, |\Phi|^{2}) =
\left. \Omega(v, \mu, T, \Phi(x), \Phi^{\ast}(x))
\right|_{\Phi, \Phi^{\ast} = \: \mbox{\tiny const}}
                                    \label{Tree.Potential}
\end{equation}
and the one-loop (quantum) correction
\begin{equation}
\Omega^{(1)}(v, \mu, T, |\Phi|^{2}) =
\frac{T}{2} \mbox{Tr Ln} \Gamma_+^{-1} +
\frac{T}{2} \mbox{Tr Ln} \Gamma_-^{-1}    \label{Quantum.Potential}
\end{equation}
Here the Green functions
$\Gamma_\pm$ are given by
\begin{eqnarray}
&& \Gamma_\pm^{-1}(\tau, \mbox{\bf r})  =
\nonumber           \\
&&
\left[ \frac{\beta \delta^{2} \Omega}
{\delta \Phi^{\ast}(\tau, \mbox{\bf r}) \delta \Phi(0,0)}
\pm \frac{\beta \delta^{2} \Omega}
{\delta \Phi^{\ast}(\tau, \mbox{\bf r}) \delta \Phi^{\ast}(0,0)}
\right]_{\Phi=\Phi^{\ast}=|\Phi| = \: \mbox{\tiny const}}
\nonumber            \\
&& = \frac{1}{V} \delta(\tau) \delta(\mbox{\bf r}) +
\mbox{tr} [G(\tau, \mbox{\bf r}) \tau_{+} G(-\tau, -\mbox{\bf r}) \tau_{-}]
\nonumber            \\
&& \pm
\mbox{tr} [G(\tau, \mbox{\bf r}) \tau_{-} G(-\tau, -\mbox{\bf r}) \tau_{-}].
                            \label{Green.fluct}
\end{eqnarray}
Strictly speaking one can only use this factorization into
$\Gamma_+^{-1} \Gamma_-^{-1}$ in the limit of small momenta and
frequencies and high carrier density \cite{Popov}. While the
representation (\ref{statsum}), (\ref{Effective.Action}) is
convenient for the studying of the LRO transition, the formation of
the BKT and pseudogap phases demands a more subtle treatment. This
is related to the fact that these phases do not display LRO. This
is evident in the original 2D model \cite{Gusynin} where LRO is
forbidden by the 2D theorems \cite{Coleman}.

Thus one must avoid the presupposition about the existence of LRO
used to write down the expression (\ref{Effective.Potential}).
The illegal step in the treatment based on representation
(\ref{statsum}), (\ref{Effective.Action}) is the fixing of a
definite value to the phase $\theta$ of the complex order parameter
$\Phi$ which is performed when one obtains Eq. (\ref{Effective.Potential}).
To avoid this dangerous step one must use the modulus, $\rho(x)$, and
phase, $\theta(x)$, parameterization of the order parameter
$\Phi(x) = \rho(x) \exp[i \theta(x)]$ as first pointed out by
Witten \cite{Witten}.
This particular choice of parameterisation ensures that $\Phi(x)$ is
single-valued with period $2 \pi$.
At the same time as this replacement by modulus-phase variables,
one must reparameterize the Nambu spinor as
$\psi_{\sigma}(x) = \chi_{\sigma}(x) \exp[i \theta(x)/2]$.

As a result we have obtained \cite{QSh} instead of (\ref{statsum}),
(\ref{Effective.Action}) the following representation
\begin{equation}
Z(v, \mu, T) = \int \rho {\cal D} \rho {\cal D} \theta
\exp{[-\beta \Omega (v, \mu, T, \rho(x), \partial \theta (x))]},
                                    \label{statsum.phase}
\end{equation}
where the one-loop effective action $\Omega (v, \mu, T, \rho(x), \partial
\theta (x))$ now depends on the  modulus-phase variables and has been
evaluated in \cite{QSh}
where we have shown that
\begin{equation}
\Omega \simeq
\Omega _{kin} (v, \mu, T, \rho, \partial \theta) +
\Omega _{pot}^{\mbox{\tiny MF}} (v, \mu, T, \rho)
                  \label{kinetic.phase+potential}
\end{equation}
where the potential energy term
in terms of $\rho^2$ is identical
to the mean-field potential in the BCS
approximation but with $|\Phi|^2$ replaced by $\rho^2$.  Thus
$T_\rho$, the temperature at which $\rho = 0$, is in this
approximation identical to the BCS mean field temperature
$T^{MF}_c$. However, although $T_\rho = T^{MF}_c$ in the mean-field
approximation for $\rho(x)$, the two temperatures have a very
different basis, both mathematically and physically \cite{Gusynin,QSh}.

It has been shown that, in the model under consideration, the
kinetic energy term reduces in lowest order to \cite{QSh}
the Hamiltonian for the
classical spin quasi-2D XY model \cite{Hikami} (see also
\cite{Ramakrishnan,Ichinose}) and we have simply used their expression
for the BKT transition in the highly anisotropic case, when the vortex
ring excitations are irrelevant \cite{Hikami}.

A self-consistent calculation of $T_{\rm BKT}^q$ as  a function of
$n_{f}$ also requires additional equations for $\rho$ and $\mu$.
The relevant equations are given by minimizing the potential
$\Omega_pot$ with respect to $\rho$ and fixing the number density.

          The numerical investigation of $T_{\rm BKT}^q$ and
$T_\rho$ has been carried out (see \cite{QSh}) and fortunately gives
results not too different from the 2D case \cite{Gusynin}.

\section{Superconducting transition into the phase with long-range order}
\setcounter{equation}{0}

As discussed in Section~II to study CLRO formation one should
use the effective potential (\ref{Effective.Potential}) obtained
in the Gaussian approximation.

There are two limiting cases where one can obtain an analytical
solution, namely, the low and high density limits. For small
concentrations of the carriers we have previously obtained the critical
temperature \cite{GLSh.PhysicaC} using the approach developed by
Nozi\`eres and Schmitt-Rink \cite{Nozieres} (see also
\cite{Randeria.3D,Haussmann}), while for the high density case we
employ a new method similar to that used in \cite{Ichinose}.

The main point of \cite{Nozieres} is the solution of the system of
number and gap equations derived using the potential
(\ref{Effective.Potential}) but with the one-loop corrections
(\ref{Quantum.Potential}) taken on the critical line
$\Phi = \Phi^{\ast} = 0$. This means that the effect of the
fluctuations is only included through the number equation,
while the gap equation is derived using the mean field potential.
Thus one
always recovers the standard mean-field (BCS) gap equation
\begin{equation}
\frac{1}{V} = \int \frac{d \mbox{\bf k}}{(2 \pi)^{3}}
\frac{1}{2 \xi(\mbox{\bf k})}
\tanh{\frac{\xi(\mbox{\bf k)}}{2 T_{c}}},              \label{gap}
\end{equation}
where  $\xi(\mbox{\bf k)} = k_\perp^2 / 2m - (m_z d^2)^{-1}
\cos k_z d - \mu$, while the
corresponding number equation has the form
\begin{equation}
n_{F} (\mu, T_c)  +  2 n_{B} (\mu, T_c) = n_{f}.
                                       \label{numberBoson}
\end{equation}
One can see from (\ref{numberBoson}) that the fermions are divided
into two coexisting systems: fermi-particles, or
unbound fermions with density $n_{F}(\mu, T_{c})$, and local pairs,
or bosons with density $n_{B}(\mu, T_{c})$.
In the extreme Bose limit $n_F=0$ the temperature has already been
shown to be the Bose condensation temperature of an ideal quasi-2D
Bose-gas \cite{GLSh.PhysicaC,Wen}

We now consider the high-density limit. This case is characterized
by the condition that the Fermi surface is not disturbed by the
attractive interaction, i.e. the contribution of the bosons to
(\ref{numberBoson}) is negligible and $\mu \simeq \epsilon_{F}$.
Thus one need only study the gap equation. It is evident that the
Nozi\`eres and Schmitt-Rink approach which yields the mean-field
equation (\ref{gap}) cannot describe a quasi-2D system adequately.
Indeed it follows from the 2D theorems
that in the limit $m_{z} \to \infty$, when the system becomes two
dimensional, the value of $T_{c}$ must go to zero. Clearly this can
never be obtained from Eq. (\ref{gap}).

To obtain the gap equation which does describe the quasi-2D system
one must include the fluctuations. This can be done if uses the effective
potential (\ref{Effective.Potential}) with the one-loop correction
(\ref{Quantum.Potential}) but without setting $\Phi = \Phi^{\ast} =0$
prior to taking the derivative with respect to $\Phi$.

Thus the gap equation takes the form
\begin{eqnarray}
&& \left. \frac{\partial
\Omega_{pot}^{\mbox{\tiny MF}}(v, \mu, T_{c}, |\Phi|^{2})}
{\partial |\Phi|^{2}} \right|_{\Phi = \Phi^{\ast} = 0} \nonumber \\
&& +  \left. \frac{\partial
\Omega^{(1)}(v, \mu, T_{c}, |\Phi|^{2})}
{\partial |\Phi|^{2}} \right|_{\Phi = \Phi^{\ast} = 0} = 0
                         \label{Bgap.def}
\end{eqnarray}
where the quantum correction (\ref{Quantum.Potential}) which was
omitted in (\ref{gap}) has now been included.

There is a subtle point related to Eq.(\ref{Bgap.def}).
The value of $T_{c}$ defined by Eq.(\ref{Bgap.def})
should be less then the mean-field temperature, $T_{c}^{MF}$
defined by Eq.(\ref{gap}) (and given by (\ref{Tc.mean})) in
the high-density limit $\mu = \epsilon_{F} \gg T_{c}^{MF}$.
However sketching the tree-potential $\Omega_{pot}^{MF}(T)$
as a function of $\Phi$ for $T < T_{c}^{MF}$ yields the Mexican hat
shape. This has a minimum at $|\Phi|^2 = \Phi^2_{min} \ne 0$
(where $\Phi_{min}$ is simply the mean-field value for $\Phi$) but a
maximum at $\Phi = \Phi^{\ast} =0$.
Thus the role of the correction to the effective potential in
Eq.(\ref{Bgap.def}) is to transform the maximum of the tree-potential
to a minimum of the full potential $\Omega_{pot}(|\Phi|^{2})$.

Unfortunately, the one-loop correction $\Omega^{(1)}$ defined by
(\ref{Quantum.Potential}) is ill-defined (complex) at the point of
interest. The temperature at which the potential just becomes complex
in fact gives the Thouless criteria of superconductivity in BCS theory.
However, in our treatment this is only an indication that the
one-loop approximation fails at the point $\Phi = \Phi^{\ast} = 0$
and is evidently related to the non-convexity of
$\Omega_{pot}^{MF}(|\Phi|^{2})$ at this point.

This situation is standard in quantum field theory
if one considers the class of theories with tree-level symmetry
breaking and an extensive literature exists
\cite{Miransky,Stevenson,EWeinberg2,Sher}. In our case to describe
a homogeneous state where $|\Phi|^2$ is uniform one should
replace $\Omega_{pot}$ by the so called Gaussian \cite{Stevenson} or
modified \cite{EWeinberg2} effective potential which coincides with
$\Omega_{pot}$ in the region where the latter is well-defined.

One could now find the modified potential, but we will
use here, in our opinion, a more transparent consideration which will
allow us to evaluate the one-loop correction not at $\Phi=0$ but
in the region where it is well-defined and coincides with
the modified potential. Our results can then be straightforwardly
related to the modified potential of Weinberg \cite{EWeinberg2}.

Let us assume that $T_{c} \ltwid T_{c}^{MF}$ which means the point
$\Phi_{min}$ is close to zero. At this point $\Omega_{pot}^{MF}$ is
surely convex and the one-loop correction (\ref{Quantum.Potential})
is real and well-defined. Thus for $T_{c} \ltwid T_{c}^{MF}$ one can
approximate Eq.(\ref{Bgap.def}) by
\begin{eqnarray}
&& \frac{1}{v} \left.
\frac{\partial \Omega_{pot}^{MF}(v, \epsilon_{F},T_{c}, \Phi^{2})}
{\partial \Phi^{2}}  \right|_{\Phi = 0} +
\nonumber          \\
&& \frac{1}{v} \left.
\frac{\partial \Omega^{(1)}(v, \epsilon_{F},T_{c}, \Phi^{2})}
{\partial \Phi^{2}}  \right|_{\Phi = \Phi_{min}} = 0 \,,
                             \label{Bgap.approx}
\end{eqnarray}
where we choose $\Phi$ to be real and the value of $\Phi_{min}$
is simply the well-known mean-field BCS value for $\Phi$ at temperature
$T$ (see e.g. \cite{Fetter} and Appendix A Eq. (\ref{gap.depen})).

Therefore to solve the approximated gap equation (\ref{Bgap.approx})
one has to calculate
\begin{eqnarray}
&& \frac{1}{v} \left.
\frac{\partial \Omega^{(1)}(v, \epsilon_{F},T, \Phi^{2})}
{\partial \Phi^{2}}
\right|_{\Phi = \Phi_{min}} =
\frac{1}{2} \frac{T}{(2 \pi)^{3}}
\sum_{\pm} \sum_{n = -\infty}^{\infty}
\nonumber                  \\
&&
\int d \mbox{\bf K} \Gamma_{\pm} (i \Omega_{n}, \mbox{\bf K}) \left.
\frac{\partial \Gamma_{\pm}^{-1} (i \Omega_{n}, \mbox{\bf K})}
{\partial \Phi^{2}} \right|_{\Phi = \Phi_{min}}\, ,
                                       \label{derivative}
\end{eqnarray}
where ${\bf K}=({\bf K}_\perp,K_z)$ and, starting from
(\ref{Green.fluct}), one can obtain the Green's functions as
a function of $\Phi$ in the momentum representation
\begin{eqnarray}
&& \Gamma_{\pm}^{-1}(i \Omega_{n}, \mbox{\bf K}) =
\frac{1}{V}  + \frac{T}{(2 \pi)^{3}} \sum_{l = -\infty}^{\infty}
\int d \mbox{\bf k} \times
\nonumber                    \\
&& \frac{[i \omega_{l}
- \xi_{-}] [i \omega_{l} + i \Omega_{n} + \xi_{+}] \pm \Phi^{2}}
{[\omega_{l}^{2} + \xi_{-}^{2} + \Phi^{2}]
[(\omega_{l} + \Omega_{n})^{2} + \xi_{+}^{2} + \Phi^{2}]},
                           \label{Gamma(pm)}
\end{eqnarray}
where we have introduced the short-hand notation
$\xi_{\pm} = \xi(\mbox{\bf k} \pm \mbox{\bf K}/2)$ and
$\Omega_{n} = 2 \pi n T$ , $\omega_{l} = \pi (2l + 1) T$ are odd and
even Matsubara frequencies, respectively.

Since $\Gamma^{-1}_-(0,{\bf 0}) = 0$ is simply the BCS gap equation
it has solution $\Phi$ equal to the BCS value $\Phi_{min}$ i.e. for
$\Phi=\Phi_{min}$, $\Gamma^{-1}_-(0,\mbox{\bf K})$ has a zero at
$\mbox{\bf K} = 0$. This gives rise to the only singular term in
(\ref{derivative}) for $\Phi = \Phi_{min}$,
namely the pole in $\Gamma_{-}(0, \mbox{\bf K})$ at $\mbox{\bf K} = 0$.
One can therefore write
\begin{eqnarray}
&& \frac{1}{v} \left.
\frac{\partial \Omega^{(1)}(v, \epsilon_{F},T, \Phi^{2})}
{\partial \Phi^{2}}
\right|_{\Phi = \Phi_{min}}  \simeq
\nonumber                \\
&& \frac{1}{2} \frac{T}{(2 \pi)^{3}} \int d \mbox{\bf K} \left.
\Gamma_{-} (0, \mbox{\bf K}) \right|_{\Phi=\Phi_{min}}
\left.  \frac{\partial \Gamma_{-}^{-1} (0, \mbox{\bf K})}
{\partial \Phi^{2}}
\right|_{\Phi = \Phi_{min}}
                     \label{derivative.approx}
\end{eqnarray}

In order to perform the calculations analytically we use the derivative
expansion for the Green function $\Gamma^{-1}_{-}$,
\begin{equation}
\left. \Gamma_{-}^{-1} (0, \mbox{\bf K}) \right|_{\Phi=\Phi_{min}} =
\frac{m_{\perp}}{2 \pi d}
\left[ a\mbox{K}_{\perp}^{2} + b [1 - \cos{K_z d}] \right] \, ,
                    \label{Gamma(-).der}
\end{equation}
(see Appendix~A, Eqs.(\ref{Gamma.normal.der}) and (\ref{ab}))
and its derivative
\begin{equation}
\left. \frac{\partial \Gamma_-^{-1} (0, \mbox{\bf K})}
{\partial \Phi^2} \right|_{\Phi=\Phi_{min}} =
\frac{m_{\perp} c}{2 \pi d}
\end{equation}
(see Appendix~A, Eqs. (\ref{Gamma.sum}) and (\ref{c})).

Substituting these two expressions into (\ref{derivative.approx}),
one arrives at the following approximation
\begin{eqnarray}
&& \frac{1}{v} \left.
\frac{\partial \Omega^{(1)}(v, \epsilon_{F},T, \Phi^{2})}
{\partial \Phi^{2}}
\right|_{\Phi = \Phi_{min}}  \simeq
\nonumber              \\
&& \frac{1}{2} \frac{T}{(2 \pi)^{3}}
\int d \mbox{\bf K} \frac{c}
{a \mbox{\bf K}_{\perp}^{2} + b [1 - \cos{K_z d}]}.
                                     \label{derivative.approx.1}
\end{eqnarray}
One can see that Eq.(\ref{derivative.approx.1}) has no infrared
divergencies due to the presence of the third
direction ($b \neq 0$). In two dimensions it would be infrared
divergent as required by the 2D theorems \cite{Coleman}.
This equation also has an artificial ultraviolet
divergence as a result of the replacement
of the Green's function $\Gamma^{-1}$ by its derivative approximation.
Thus one should introduce a rather natural ultraviolet cutoff
$(\mbox{\bf K}_{\perp}^{max})^{2} = 2 m_{\perp} \Phi(T=0) =
2 m_{\perp} \sqrt{2 |\varepsilon_{b}| \epsilon_{F}}$
and integrate over the momentum $\mbox{\bf K}$ to
obtain the expression
\begin{equation}
\frac{1}{v} \left.
\frac{\partial \Omega^{(1)}(v, \epsilon_{F},T, \Phi^{2})}
{\partial \Phi^{2}}
\right|_{\Phi = \Phi_{min}}  \simeq
\frac{m_{\perp}}{2 \pi d} \frac{T}{2 \epsilon_{F}} |\ln \kappa| \, ,
                                     \label{correction.final}
\end{equation}
where
\begin{equation}
\kappa = \frac{1}{4 \sqrt{2}} \frac{w^2}{\epsilon_{F}^{2}}
\sqrt{\frac{\epsilon_{F}}{|\varepsilon_{b}|}}, \qquad
w = \frac{1}{m_z d^2}.
                                       \label{kappa}
\end{equation}
Substituting (\ref{correction.final}) into (\ref{Bgap.approx})
one obtains the final transcendental equation for $T_{c}$
\begin{equation}
\ln \frac{T_c}{T_{c}^{MF}} + \frac{T_c}{2 \epsilon_{F}}
|\ln \kappa| = 0 \, ,
              \label{Tc.eq}
\end{equation}
which may be rewritten in the following more convenient form
\begin{equation}
T_{c} = 2 \epsilon_{F} \frac{|\ln(T_{c}/T_{c}^{MF})|}{|\ln \kappa|}.
                       \label{Tc}
\end{equation}
Strictly speaking the equation (\ref{Tc}) is only valid when
$T_c \ltwid T_{c}^{MF}$. However, we intend to use Eq.(\ref{Tc}) even
when $T_{c} < T_{c}^{MF}$, which will allow to qualitatively
sketch the whole phase diagram of the system.

If one solves (\ref{Tc}) numerically for reasonable width one
obtains roughly linear behaviour $T_c \sim \epsilon_F \sim n_f$ for
a wide range of intermediate densities (see below). This dependence
is observed experimentally \cite{Uemura} for HTSCs samples which
have a Fermi surface.

A second feature of (\ref{Tc}) is that increasing
$|\varepsilon_{b}|$ increases both $T_{c}^{MF}$ and $T_{c}$.
This is opposite to the behaviour in the low density limit. Finally
$T_{c}$ goes to zero as $m_{z} \to \infty$ ($w \to 0$) as it must.
This is very different to the limiting behaviour of $T^q_{\rm BKT}$,
namely $T^q_{\rm BKT} \rightarrow T_{\rm BKT}$ in the
limit $m_z \to \infty$.

As stated above one can also understand the
approximation used in (\ref{Bgap.approx}) in terms of the modified
effective potential defined in \cite{EWeinberg2}. The modified
effective potential in \cite{EWeinberg2} is defined as the minimum
value for $\Omega$ given a homogeneous state where $|\Phi|^2$ is
uniform. The real part of this modified potential has the following
form
\begin{eqnarray}
&& \tilde \Omega^{(1)}(v, \mu, T, |\Phi|^{2}) =
\nonumber             \\
&& \frac{T}{2 (2 \pi)^{3}} \sum_{\pm} \sum_{n = -\infty}^{\infty}
\int_{\cal D}
d \mbox{\bf K} \ln \Gamma_{\pm}^{-1} (i \Omega_{n}, \mbox{\bf K}) \, ,
             \label{Quantum.Potential.modified}
\end{eqnarray}
where the area $\cal D$ of integration in the momentum space
includes only positive modes. One can see that
(\ref{Quantum.Potential.modified}) indeed coincides with
(\ref{Quantum.Potential}) when $\Omega^{(1)}$ is well-defined.
Furthermore the modified potential (\ref{Quantum.Potential.modified})
leads to the gap equation (\ref{Bgap.approx}) which was considered
above as the approximated one.

In the region $\Phi < \Phi_{min}$ the modified effective potential
considered above differs from the traditional effective potential,
$\Omega_{eff}(\Phi)$, which is defined as the minimum value for $\Omega$
such that the space average of $\Phi(x)$ is given by $\Phi$. It can be
shown that the conventional effective potential is in fact the convex
envelope of the modified effective potential and is real and convex
everywhere. However for $\Phi < \Phi_{min}$ it describes an inhomogeneous
mixed state where the value of $\Phi(x)$ is not uniform in space. One can
readily understand that the modified and not the original potential is
relevant for the superconducting state.

There is, however, the difference between our interpretation of the
modified potential and that of \cite{EWeinberg2}. In \cite{EWeinberg2}
the homogeneous state described by (\ref{Quantum.Potential.modified})
is considered as decaying and the rate of the decay is related to
negative modes of (\ref{Quantum.Potential}) which are not included in
(\ref{Quantum.Potential.modified}). It is physically obvious that
there is no real decay of the homogeneous superconducting state with
$\Phi < \Phi_{min}$ for $T < T_{c}$ although we have not been able
to prove this rigorously. The absence of decay is in agreement with the
interpretation of \cite{Stevenson} although it should be stressed that
the modified potential discussed here is not identical to the
Gaussian effective potential in \cite{Stevenson}.

\section{The phase diagram of the system}
\setcounter{equation}{0}

We have now obtained solutions for $T_{c}$,
$T_{\rm BKT}^{q}$ and $T_\rho$ and can build the phase diagram for the
quasi-2D system. This diagram comprises the normal phase, the pseudogap
phase, the superconducting BKT region, and the CLRO phase
with superconductivity in the bulk in order of decreasing temperature.
The practical observation of these regions (i.e. the temperature range for
which they are present) is however critically dependent on the carrier
density.

$\imath)$ We first reiterate results at very low carrier densities
$\tilde{\epsilon} \ll 1$ i.e. for a Bose liquid of isolated local pairs.

Firstly, the critical temperature is linear in $\epsilon_F$
$T_c \sim \epsilon _{F}$ (or $T_c \sim n_{f}$ as expected in the 2D case)
\cite{GLSh.PhysicaC}. For large anisotropy the proportionality constant
depends only on the particle density and the width in the z-direction.
This constant is less than $1/8$ for half-bandwidth
\begin{equation}
w \le \frac{\sqrt{2 |\epsilon_b| \epsilon_F}}{2 \sqrt{2} \exp{2}} =
\frac{\Phi(T=0)}{2 \sqrt{2} \exp{2}}
\label{bandw}
\end{equation}
where $\Phi(T=0)$ is the zero-temperature energy gap. Recall that in
the 3D case $T_c \sim n_{f}^{2/3}$ \cite{Randeria.3D}. Secondly,
contrary to the case of a 3D superconductor where $T_c$ is independent
of $V$ \cite{Randeria.3D}, in a quasi-2D system $T_c$ does depend on $V$.
In particular $T_c$ decreases with the growth of $V$.  The reason for
this is the dynamical increasing of the composite boson mass in the
third direction. Thus, the growth of $|\varepsilon _{b}|$
(or equivalently of $V$) "makes" the system more and more two-dimensional
in this model of a quasi-2D metal with a local four-fermion interaction.
It is interesting to note that a decreasing $T_c$ can also take place
in the case when the local pairs are bipolarons \cite{Alexandrov}.

Furthermore the temperature $T_{\rm BKT}^q$ always lies above $T_c$
in this limit for reasonable bandwidths (see Eq. (\ref{bandw}) and
strongly anisotropic systems \cite{Gusynin,QSh} since
\begin{equation}
T_{\rm BKT}^q > T_{\rm BKT} = \frac{\epsilon_F}{8}.
\end{equation}
In addition the BKT region and LRO order phase are of roughly equal size
so one expects to see an extensive region where the interlayer tunneling
is insufficient to produce CLRO.

This range of carrier densities is not however experimentally
important because the anomalous behaviour of HTSCs is observed when
the Fermi surface is still present.

$\imath \imath)$  We next present results for high carrier densities
($\tilde{\epsilon} \gg 10^3$).

At high carrier densities $T_c$ approaches $T_c^{MF}$ asymptotically so that
the condition $T_c \ltwid T_c^{MF}$ is satisfied and equation (\ref{Tc})
for the critical temperature is indeed valid. Furthermore  one can expand the
logarithm in equation (\ref{Tc}) to give
\begin{equation}
T_{c} = T_\rho \left(1 -
\frac{T_\rho |\ln\kappa|}{2 \epsilon_{F}} \right),\qquad \tilde{\epsilon} \rtwid
10^3
                  \label{Tc.near}
\end{equation}
where $T_c^{MF}$ has been replaced by the temperature $T_\rho$ (which is
identical in the approximation used in this paper).

For these carrier densities one also finds that \\
$\rho(T_{\rm BKT})/T_{\rm BKT} \ll 1$ which yields
the following estimation for the BKT transition temperature
\begin{equation}
T_\rho > T_{\rm BKT}^{q} > T_{\rm BKT} \simeq T_{\rho}
\left( 1 - \frac{4 T_{\rho}}{\epsilon_{F}}\right), \qquad
\tilde{\epsilon} \rtwid 10^3.
                              \label{BKT.asymptot}
\end{equation}
From (\ref{Tc.near}) and (\ref{BKT.asymptot}) it follows that
$T_{\rm BKT}^q \ge T_c$ for reasonable anisotropies, bandwidths
satisfying (\ref{bandw}) and
for particle density, $\tilde{\epsilon} > 1/2 (e/2)^8$
which is clearly satisfied in the high-density region. Furthermore
both $T_c$ and $T_{\rm BKT}^q$ tend  asymptotically to $T_\rho = T_c^{MF}$
in the high density limit. Thus both the pseudogap and BKT regions
vanish asymptotically in this limit and one recovers the standard BCS
behaviour with a transition directly into a superconducting state with CLRO
at the BCS critical temperature as obtained by \cite{Kats} in the quasi-1D
case and in the weak-coupling limit in 2D
\cite{Randeria.1997,Gusynin,Babaev}.

$\imath \imath \imath)$ The physical interest lies at intermediate
densities ($\tilde{\epsilon} \sim 10-10^3$). Optimal doping corresponds
to $\tilde{\epsilon} \sim 3 \cdot 10^2 - 10^3$ \cite{Carter} and we
show this range of densities in Fig.~1. The precise value of
$\tilde{\epsilon}$ for optimal doping is however strongly dependent
on the anisotropy of the system.  For these relatively high densities
the condition  $T_{\rm BKT}^q, T_c \ltwid T_c^{MF}$ is still satisfied
and equation (\ref{Tc}) is a good approximation. For example, for
$\tilde{\epsilon} = 700,$ $(T_\rho-T_c)/T_c \approx 0.15$, indicating
that both the pseudogap and BKT regions are small.  One can see from
Eq. (\ref{Tc.near}) that
\begin{equation}
 \frac{T_\rho - T_c}{T_c} \rightarrow 0, \qquad
\tilde{\epsilon} \rightarrow \infty
\end{equation}
This gives an asymptotic disappearance of the pseudogap since the
fraction of the temperature range in the pseudogap phase goes to
zero. This corresponds to the experimental observations.
Furthermore one finds that $T_c$ is very close to $T^q_{\rm BKT}$
which is in agreement with the experimental observation of
at most a very narrow crossover region in temperature from two-dimensional to
three-dimensional superconductivity reported in \cite{Stamp,Zverev} for
optimally doped cuprates.

Due to the fact that the Fermi surface is still present for the
densities where the pseudogap develops, one might expect the
pseudogap phenomena to occur at  $\tilde{\epsilon} \sim 10-10^2$.
Extrapolating (\ref{Tc}) to these densities one finds that $T_{c}$
is always less then $T_{\rm BKT}^{q}$. More essentially $(T_{\rho}-
T_{\rm BKT}^{q})/T_{\rho} \sim 1$, i.e. the pseudogap phase is big
enough to make it observed and sufficient to explain the observed
anomalies in the underdoped compounds. On the other hand $(T_{\rm
BKT}^q - T_c)/T_c \ll 1$ so that the BKT region again remains
relatively small. In this density range these two temperatures are
so close that one cannot see the difference on the phase diagram. The phase
diagram is thus unchanged from that shown in \cite{QSh} since the lines
for $T_c$ and $T^q_{\rm BKT}$ coincide.  Thus for all carrier densities
appropriate to HTSCs the superconducting transition is practically directly
to the phase with CLRO.  Since the temperature range for the BKT region is
only a few degrees it has until recently \cite{Zverev} lain within the
observed critical region.

We note that our results for $T_c$ imply that the quasi-2D BKT region is
decreasing in size with decreasing density. We believe that this is not
the case physically particularly since it has been shown that
in the Bose limit the quasi-2D BKT region is large.  Thus we believe that
this is an artifact of our approximation since it becomes less accurate
as $T_c$ moves away from $T_\rho$ which occurs with decreasing density (see
discussion after Eq. (\ref{Tc})).  Thus, although one can say that this
transitional temperature range is small, our approximation is not adequate to
describe its dependence on the carrier density.

\section{Concluding Remarks}
\setcounter{equation}{0}

We have derived a phase diagram for a superconducting quasi-2D
system in which one has four regions. These are the normal phase, the
pseudogap region, the BKT region and the phase with conventional bulk
superconductivity (CLRO).

For all realistic carrier densities we find the BKT
region to be only a few degrees. This implies that it lies within the
critical region and is difficult to observe experimentally. In fact
$T_{\rm BKT}^q$ is slightly overestimated (since for example the
effect of vortex rings is omitted). In contrast the value of $T_c$ is
underestimated by the approximation used since only one loop corrections
have been included \cite{Ichinose}. Thus the BKT region, if present, is
even smaller than calculated.

In contrast the pseudogap region is substantial for the underdoped
region where the unusual superconducting condensate formation is
responsible for the observed anomalies in underdoped HTSCs. At
optimal doping the pseudogap phase is also negligible and one
recovers the standard BCS behaviour.

With the calculation of $T_c$ we are able to present a picture in
which two dimensional behaviour predominates in the pseudogap phase
but where the superconducting phase is three dimensional in line with the
recent result of \cite{Vilk}.

\section*{Acknowledgments}
We gratefully acknowledge E.V.~Gorbar and V.P.~Gusynin for proofreading the
manuscript and valuable suggestions. We also thank N.J.~Davidson,
V.A.~Miransky, I.A.~Shovkovy and O.~Tchernyshyov for fruitful discussions.
One of us (S.G.Sh) is grateful to the members of the
Department of Physics of the University of Pretoria for hospitality.
R.M.Q and S.G.Sh acknowledge the financial support of the Foundation for
Research Development, Pretoria.

\renewcommand{\theequation}{\Alph{section}.\arabic{equation}}
\appendix

\section{The calculation of the Green functions $\Gamma_{\pm}$}
\setcounter{equation}{0}

We derive here the Green functions
$\Gamma_{\pm}(i \Omega_{n}, \mbox{\bf K})$ in the derivative approximation
for the large density ($\mu \simeq \epsilon_{F} \gg T$) limit starting from
expression (\ref{Gamma(pm)}). Since $\Phi/T \ll 1$ for $T \ltwid T_{c}$
one can use the "high-temperature" approximation
\begin{equation}
\frac{1}{\omega_{n}^{2} + \xi^{2} + \Phi^{2}} \simeq
\frac{1}{\omega_{n}^{2} + \xi^{2}} -
\frac{\Phi^{2}}{[\omega_{n}^{2} + \xi^{2}]^{2}},
                      \label{approximation}
\end{equation}
which gives
\begin{eqnarray}
&& \Gamma_{\pm}^{-1}(i \Omega_{n}, \mbox{\bf K}) =  \frac{1}{V}  +
T \sum_{l = -\infty}^{\infty} \int \frac{d \mbox{\bf k}}{(2 \pi)^{3}} \times
\nonumber               \\
&& \left\{
\frac{1}{[i \omega_{l} + \xi_{-}][i \omega_{l} + i \Omega_{n} - \xi_{+}]}
+ \frac{2 \Phi^{2} \pm \Phi^{2}}{[\omega_{l}^{2} + \xi^{2}]^{2}} \right\}.
                           \label{Gamma.high}
\end{eqnarray}
where to lowest order one can set $\Omega_n={\bf K}=0$ in the $\Phi^2$ term.
Performing the summation over fermion Matsubara frequencies one
arrives at the following result
\begin{equation}
\Gamma_{\pm}^{-1}(i \Omega_{n}, \mbox{\bf K}) =
\Gamma^{-1}(i \Omega_{n}, \mbox{\bf K}) +
\frac{m_{\perp}}{2 \pi d} c ( 2 \Phi^{2} \pm \Phi^{2}),
                           \label{Gamma.sum}
\end{equation}
where in for large densities
\begin{equation}
c = \frac{7 \zeta(3)}{8 \pi^{2} T^{2}}
                           \label{c}
\end{equation}
and $\Gamma(i \Omega_{n}, \mbox{\bf K})$ is the Green function
of the order parameter fluctuations in the normal ($\Phi, \Phi^{\ast} =0$)
state:
\begin{eqnarray}
\Gamma^{-1}(i \Omega_{n}, \mbox{\bf K}) = && \frac{1}{V}  -
\frac{1}{2} \int \frac{d \mbox{\bf k}}{(2 \pi)^{3}}
\frac{1}{\xi_{+} + \xi_{-} - i \Omega_{n}}
\nonumber           \\
&& \times   \left[
\tanh \frac{\xi_{+}}{2 T} +  \tanh \frac{\xi_{-}}{2 T} \right].
                           \label{Gamma.normal.def}
\end{eqnarray}
The 2D limit ($w=0$) of the Green function was studied by Shovkovy et al.
\cite{Shovkovy,LSh.FNT.1997} and we need to generalize the expression
to the quasi-2D case.
Substituting the coupling constant $V$ expressed via
the bound state energy (\ref{bound.energy}) into
(\ref{Gamma.normal.def}) and performing simple
transformations on the hyperbolic functions one arrives at
\begin{eqnarray}
&& \Gamma^{-1}(i \Omega_{n}, \mbox{\bf K}) =
\frac{m_{\perp}}{4 \pi d} \ln \frac{2 W}{|\varepsilon_{b}|}
\nonumber               \\
&& - \frac{m_{\perp}}{8 \pi^{2} d} \int_{0}^{2W} dx \int_{0}^{2 \pi} dt
\frac{1}{x + s(t, K_{\perp}, K_{z}) - i \Omega_{n}}
\nonumber               \\
&& \times
\tanh \frac{x + s(t, K_{\perp}, K_{z})}{4 T}
\nonumber               \\
&& + \frac{m_{\perp}}{8 \pi^{2} d} \int_{0}^{2W} dx \int_{0}^{2 \pi} dt
\frac{2}{\pi} \int_{0}^{\pi/2} d \varphi
\frac{1}{x + s(t, K_{\perp}, K_{z}) - i \Omega_{n}}
\nonumber               \\
&& \times
\tanh \frac{x + s(t, K_{\perp}, K_{z})}{4 T} \times
\nonumber           \\
&& \frac{ \cosh\frac{\ds d(t, K_{\perp}, K_{z}, \cos \varphi)}{\ds 2T}
-1}
{\cosh \frac{\ds x + s(t, K_{\perp}, K_{z})}{\ds 2T}  +
\cosh \frac{\ds d(t, K_{\perp}, K_{z}, \cos \varphi )}{\ds 2T} },
                \label{Gamma.normal.1}
\end{eqnarray}
where we introduced the short-hand notations
$$
s(t, K_{\perp}, K_{z}) = \frac{K_{\perp}^{2}}{4 m_{\perp}}
- 2 w \cos t \cos \left( \frac{K_{z} d}{2} \right) - 2 \mu
$$ and
$$ d(t, K_{\perp}, K_{z}, \cos \varphi) =
\frac{k_{\perp} K_{\perp} \cos \varphi }{m_{\perp}} +
2 w \sin t \sin \left( \frac{K_{z} d}{2} \right).
$$

Using the identities:
\begin{eqnarray}
&& \frac{1}{x + s(t, K_{\perp}, K_{z}) - i \Omega_{n}} =
\frac{1}{x + s(t, K_{\perp}, K_{z})}
\nonumber                  \\
&& +
\frac{i \Omega_{n}}{[x + s(t, K_{\perp}, K_{z})]
[x + s(t, K_{\perp}, K_{z}) - i \Omega_{n}]} \,,
                        \label{simple.frac}
\end{eqnarray}
and
\begin{equation}
\int_{0}^{a} \frac{dx}{x} \tanh x = \ln \frac{4 a \gamma}{\pi}
- \int_{a}^{\infty} \frac{dx}{x} (\tanh x - 1) \,,
\mbox{at} \quad a > 0 \, ,
                                \label{integral}
\end{equation}
($\ln \gamma$ is the Euler constant), one can take the limit
$W \to \infty$ in (\ref{Gamma.normal.1}) and come to the
expression:
\begin{eqnarray}
&& \Gamma^{-1}(i \Omega_{n}, \mbox{\bf K}) =
\frac{m_{\perp}}{4 \pi d} \ln \frac{\pi T}{|\varepsilon_{b}| \gamma}
\nonumber               \\
&& + \frac{m_{\perp}}{8 \pi^{2} d} \int_{0}^{2 \pi} dt \int_{0}^{1}
\frac{d x}{x} \tanh \left[x \frac{s(t, K_{\perp}, K_{z})}{4T} \right]
\nonumber               \\
&& - i \Omega_{n}
\frac{m_{\perp}}{8 \pi^{2} d} \int_{0}^{2 \pi} dt \int_{0}^{\infty} dx
\nonumber               \\
&& \frac{1}{[x + s(t, K_{\perp}, K_{z})]
[x + s(t, K_{\perp}, K_{z}) - i \Omega_{n}]}
\nonumber                \\
&& \times \tanh \frac{x + s(t, K_{\perp}, K_{z})}{4 T}
\nonumber               \\
&& + \frac{m_{\perp}}{8 \pi^{2} d} \int_{0}^{2 \pi} dt \int_{0}^{\infty} dx
\frac{1}{\pi} \int_{0}^{\pi} d \varphi
\frac{1}{x + s(t, K_{\perp}, K_{z}) - i \Omega_{n}}
\nonumber               \\
&& \times \tanh \frac{x + s(t, K_{\perp}, K_{z})}{4 T} \times
\nonumber               \\
&& \frac{\cosh \frac{\ds d(t, K_{\perp}, K_{z}, \cos \varphi)}{\ds 2T} -1}
{\cosh \frac{\ds x + s(t, K_{\perp}, K_{z})}{\ds 2T} +
\cosh \frac{\ds d(t, K_{\perp}, K_{z}, \cos \varphi )}{\ds 2T} } \,.
                                           \label{Gamma.normal.2}
\end{eqnarray}
One may easily show that in the 2D limit $w \to 0$ the
expression (\ref{Gamma.normal.2}) transforms into the Green
function obtained in \cite{Shovkovy}. For example, the last
term of (\ref{Gamma.normal.2}) gives the exact representation
for the imaginary part of this function.

We, however, are interested in the derivative expansion
of $\Gamma^{-1} (0, \mbox{\bf K})$ for the
anisotropies $w/T \ll 1$ and large densities
which has the following form
\begin{equation}
\Gamma^{-1}(0, \mbox{\bf K}) = \frac{m_{\perp}}{2 \pi d}
\left[ \ln \left( \frac{T}{T_{c}^{MF}} \right) + a\mbox{K}_{\perp}^{2}
+ b [1 - \cos{K_z d}] \right] \, ,
                      \label{Gamma.normal.der}
\end{equation}
where
\begin{equation}
T_{c}^{MF} = \frac{\gamma}{\pi}
\sqrt{2 |\varepsilon_{b}| \epsilon_{F}}
                     \label{Tc.mean}
\end{equation}
is the mean-field transition temperature obtained from
the mean-field gap equation (\ref{gap});
\begin{equation}
a = \frac{7 \zeta(3)}{(4 \pi)^2} \frac{\epsilon_{F}}{m_{\perp} T^2},
\qquad b = \frac{7 \zeta(3)}{(4 \pi)^2} \frac{w^2}{T^2}.
                                      \label{ab}
\end{equation}
Again the expression (\ref{Gamma.normal.der}) is in
the correspondence with the Green function used in
\cite{LSh.FNT.1997}. We note also that the anisotropy
$\alpha (\epsilon_{F}) = b d^2/2 a$ is equal to the anisotropy
that was found in the quasi-2D BKT model \cite{QSh}.

Finally we should stress that if one substitutes into
(\ref{Gamma.sum}) (see also (\ref{Gamma.normal.der}) the
well-known mean-field BCS dependence (see e.g. \cite{Fetter})
\begin{equation}
\Phi^{2}_{min}(T) = \frac{8 \pi^{2} (T_{c}^{MF})^{2}}{7 \zeta(3)}
\left( 1 - \frac{T}{T_{c}^{MF}} \right)
              \label{gap.depen}
\end{equation}
of the gap on the temperature near $T_{c}^{MF}$,
one obtains (\ref{Gamma(-).der}) which shows that
$\Gamma_{+}(0, \mbox{\bf K}) > 0$, while
$\Gamma_{-}(0, \mbox{K}) \leq 0$ with the pole at
$\mbox{K} =0$.


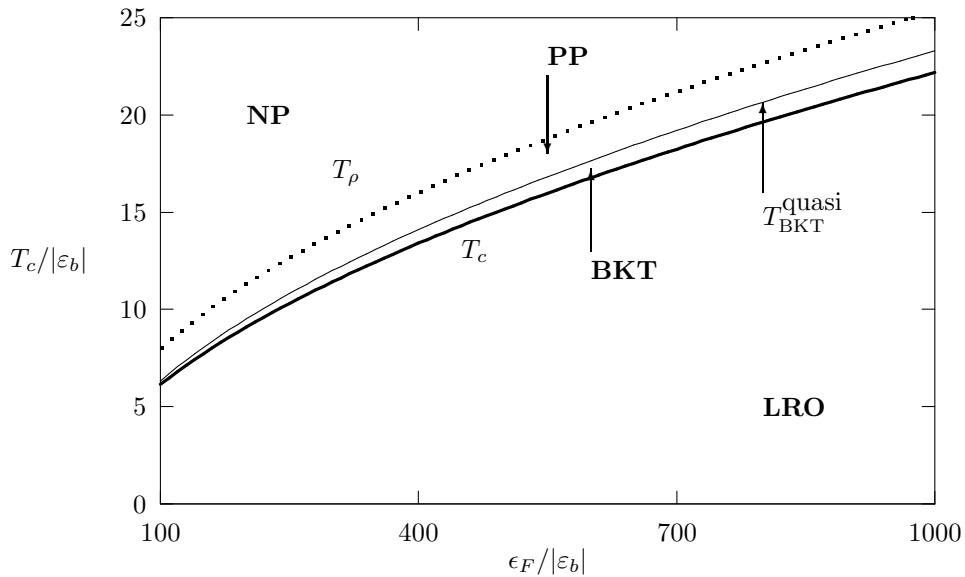
\begin{figure}

\begin{center}
\setlength{\unitlength}{0.240900pt}
\ifx\plotpoint\undefined\newsavebox{\plotpoint}\fi
\sbox{\plotpoint}{\rule[-0.200pt]{0.400pt}{0.400pt}}%
\special{em:linewidth 0.4pt}%
\begin{picture}(1500,900)(0,0)
\font\gnuplot=cmr10 at 10pt
\gnuplot
\put(220,113){\special{em:moveto}}
\put(1436,113){\special{em:lineto}}
\put(220,113){\special{em:moveto}}
\put(240,113){\special{em:lineto}}
\put(1436,113){\special{em:moveto}}
\put(1416,113){\special{em:lineto}}
\put(198,113){\makebox(0,0)[r]{0}}
\put(220,266){\special{em:moveto}}
\put(240,266){\special{em:lineto}}
\put(1436,266){\special{em:moveto}}
\put(1416,266){\special{em:lineto}}
\put(198,266){\makebox(0,0)[r]{5}}
\put(220,419){\special{em:moveto}}
\put(240,419){\special{em:lineto}}
\put(1436,419){\special{em:moveto}}
\put(1416,419){\special{em:lineto}}
\put(198,419){\makebox(0,0)[r]{10}}
\put(220,571){\special{em:moveto}}
\put(240,571){\special{em:lineto}}
\put(1436,571){\special{em:moveto}}
\put(1416,571){\special{em:lineto}}
\put(198,571){\makebox(0,0)[r]{15}}
\put(220,724){\special{em:moveto}}
\put(240,724){\special{em:lineto}}
\put(1436,724){\special{em:moveto}}
\put(1416,724){\special{em:lineto}}
\put(198,724){\makebox(0,0)[r]{20}}
\put(220,877){\special{em:moveto}}
\put(240,877){\special{em:lineto}}
\put(1436,877){\special{em:moveto}}
\put(1416,877){\special{em:lineto}}
\put(198,877){\makebox(0,0)[r]{25}}
\put(220,113){\special{em:moveto}}
\put(220,133){\special{em:lineto}}
\put(220,877){\special{em:moveto}}
\put(220,857){\special{em:lineto}}
\put(220,68){\makebox(0,0){100}}
\put(625,113){\special{em:moveto}}
\put(625,133){\special{em:lineto}}
\put(625,877){\special{em:moveto}}
\put(625,857){\special{em:lineto}}
\put(625,68){\makebox(0,0){400}}
\put(1031,113){\special{em:moveto}}
\put(1031,133){\special{em:lineto}}
\put(1031,877){\special{em:moveto}}
\put(1031,857){\special{em:lineto}}
\put(1031,68){\makebox(0,0){700}}
\put(1436,113){\special{em:moveto}}
\put(1436,133){\special{em:lineto}}
\put(1436,877){\special{em:moveto}}
\put(1436,857){\special{em:lineto}}
\put(1436,68){\makebox(0,0){1000}}
\put(220,113){\special{em:moveto}}
\put(1436,113){\special{em:lineto}}
\put(1436,877){\special{em:lineto}}
\put(220,877){\special{em:lineto}}
\put(220,113){\special{em:lineto}}
\put(45,495){\makebox(0,0){$T_c/|\varepsilon_{b}|$}}
\put(828,23){\makebox(0,0){$\epsilon_{F}/|\varepsilon_{b}|$}}
\put(1166,571){\makebox(0,0)[l]{$T_{\rm BKT}^{\mbox{quasi}}$}}
\put(490,633){\makebox(0,0)[l]{$T_{\rho}$}}
\put(693,510){\makebox(0,0)[l]{$T_{c}$}}
\put(1166,266){\makebox(0,0)[l]{${\bf LRO}$}}
\put(896,480){\makebox(0,0)[l]{${\bf BKT}$}}
\put(828,816){\makebox(0,0)[l]{${\bf PP}$}}
\put(355,724){\makebox(0,0)[l]{${\bf NP}$}}
\put(1166,602){\vector(0,1){141}}
\put(896,510){\vector(0,1){129}}
\put(828,785){\vector(0,-1){122}}
\sbox{\plotpoint}{\rule[-0.500pt]{1.000pt}{1.000pt}}%
\special{em:linewidth 1.0pt}%
\put(220,358){\usebox{\plotpoint}}
\put(220.00,358.00){\usebox{\plotpoint}}
\put(235.77,371.50){\usebox{\plotpoint}}
\put(251.75,384.73){\usebox{\plotpoint}}
\put(267.87,397.81){\usebox{\plotpoint}}
\put(284.35,410.39){\usebox{\plotpoint}}
\put(300.90,422.92){\usebox{\plotpoint}}
\multiput(301,423)(17.459,11.224){0}{\usebox{\plotpoint}}
\put(318.16,434.43){\usebox{\plotpoint}}
\put(335.01,446.51){\usebox{\plotpoint}}
\put(352.23,458.09){\usebox{\plotpoint}}
\multiput(355,460)(18.021,10.298){0}{\usebox{\plotpoint}}
\put(370.08,468.66){\usebox{\plotpoint}}
\put(387.69,479.65){\usebox{\plotpoint}}
\put(405.26,490.70){\usebox{\plotpoint}}
\multiput(409,493)(18.021,10.298){0}{\usebox{\plotpoint}}
\put(423.21,501.11){\usebox{\plotpoint}}
\put(441.57,510.79){\usebox{\plotpoint}}
\put(459.65,520.94){\usebox{\plotpoint}}
\multiput(463,523)(18.564,9.282){0}{\usebox{\plotpoint}}
\put(478.03,530.55){\usebox{\plotpoint}}
\put(496.22,540.55){\usebox{\plotpoint}}
\put(514.71,549.94){\usebox{\plotpoint}}
\multiput(517,551)(18.564,9.282){0}{\usebox{\plotpoint}}
\put(533.27,559.22){\usebox{\plotpoint}}
\put(551.87,568.37){\usebox{\plotpoint}}
\put(570.41,577.68){\usebox{\plotpoint}}
\multiput(571,578)(19.077,8.176){0}{\usebox{\plotpoint}}
\put(589.27,586.30){\usebox{\plotpoint}}
\put(607.96,595.27){\usebox{\plotpoint}}
\multiput(612,597)(18.845,8.698){0}{\usebox{\plotpoint}}
\put(626.88,603.81){\usebox{\plotpoint}}
\put(645.87,612.17){\usebox{\plotpoint}}
\put(664.88,620.52){\usebox{\plotpoint}}
\multiput(666,621)(18.845,8.698){0}{\usebox{\plotpoint}}
\put(683.79,629.05){\usebox{\plotpoint}}
\put(702.75,637.50){\usebox{\plotpoint}}
\multiput(706,639)(19.546,6.981){0}{\usebox{\plotpoint}}
\put(722.10,644.97){\usebox{\plotpoint}}
\put(741.24,652.94){\usebox{\plotpoint}}
\multiput(747,655)(18.845,8.698){0}{\usebox{\plotpoint}}
\put(760.30,661.11){\usebox{\plotpoint}}
\put(779.64,668.60){\usebox{\plotpoint}}
\put(798.91,676.25){\usebox{\plotpoint}}
\multiput(801,677)(19.372,7.451){0}{\usebox{\plotpoint}}
\put(818.24,683.82){\usebox{\plotpoint}}
\put(837.54,691.41){\usebox{\plotpoint}}
\multiput(842,693)(19.372,7.451){0}{\usebox{\plotpoint}}
\put(856.97,698.70){\usebox{\plotpoint}}
\put(876.45,705.87){\usebox{\plotpoint}}
\put(895.95,712.98){\usebox{\plotpoint}}
\multiput(896,713)(19.372,7.451){0}{\usebox{\plotpoint}}
\put(915.38,720.28){\usebox{\plotpoint}}
\put(934.82,727.54){\usebox{\plotpoint}}
\multiput(936,728)(19.546,6.981){0}{\usebox{\plotpoint}}
\put(954.31,734.66){\usebox{\plotpoint}}
\put(973.78,741.85){\usebox{\plotpoint}}
\multiput(977,743)(19.838,6.104){0}{\usebox{\plotpoint}}
\put(993.52,748.26){\usebox{\plotpoint}}
\put(1012.98,755.46){\usebox{\plotpoint}}
\multiput(1017,757)(19.957,5.702){0}{\usebox{\plotpoint}}
\put(1032.77,761.68){\usebox{\plotpoint}}
\put(1052.38,768.40){\usebox{\plotpoint}}
\multiput(1058,770)(19.372,7.451){0}{\usebox{\plotpoint}}
\put(1071.93,775.33){\usebox{\plotpoint}}
\put(1091.57,782.02){\usebox{\plotpoint}}
\put(1111.49,787.85){\usebox{\plotpoint}}
\multiput(1112,788)(19.372,7.451){0}{\usebox{\plotpoint}}
\put(1131.05,794.73){\usebox{\plotpoint}}
\put(1150.66,801.48){\usebox{\plotpoint}}
\multiput(1152,802)(19.957,5.702){0}{\usebox{\plotpoint}}
\put(1170.55,807.40){\usebox{\plotpoint}}
\put(1190.22,814.01){\usebox{\plotpoint}}
\multiput(1193,815)(19.838,6.104){0}{\usebox{\plotpoint}}
\put(1210.04,820.15){\usebox{\plotpoint}}
\put(1229.94,826.06){\usebox{\plotpoint}}
\multiput(1233,827)(19.957,5.702){0}{\usebox{\plotpoint}}
\put(1249.79,832.07){\usebox{\plotpoint}}
\put(1269.44,838.70){\usebox{\plotpoint}}
\multiput(1274,840)(19.838,6.104){0}{\usebox{\plotpoint}}
\put(1289.32,844.66){\usebox{\plotpoint}}
\put(1309.23,850.53){\usebox{\plotpoint}}
\multiput(1314,852)(19.957,5.702){0}{\usebox{\plotpoint}}
\put(1329.15,856.35){\usebox{\plotpoint}}
\put(1349.03,862.30){\usebox{\plotpoint}}
\multiput(1355,864)(19.838,6.104){0}{\usebox{\plotpoint}}
\put(1368.91,868.26){\usebox{\plotpoint}}
\put(1388.83,874.10){\usebox{\plotpoint}}
\multiput(1395,876)(20.136,5.034){0}{\usebox{\plotpoint}}
\put(1399,877){\usebox{\plotpoint}}
\sbox{\plotpoint}{\rule[-0.600pt]{1.200pt}{1.200pt}}%
\special{em:linewidth 1.2pt}%
\put(220,301){\special{em:moveto}}
\put(234,311){\special{em:lineto}}
\put(247,321){\special{em:lineto}}
\put(261,331){\special{em:lineto}}
\put(274,340){\special{em:lineto}}
\put(288,349){\special{em:lineto}}
\put(301,358){\special{em:lineto}}
\put(315,367){\special{em:lineto}}
\put(328,375){\special{em:lineto}}
\put(342,383){\special{em:lineto}}
\put(355,391){\special{em:lineto}}
\put(369,399){\special{em:lineto}}
\put(382,406){\special{em:lineto}}
\put(396,414){\special{em:lineto}}
\put(409,421){\special{em:lineto}}
\put(423,428){\special{em:lineto}}
\put(436,435){\special{em:lineto}}
\put(450,442){\special{em:lineto}}
\put(463,449){\special{em:lineto}}
\put(477,455){\special{em:lineto}}
\put(490,462){\special{em:lineto}}
\put(504,468){\special{em:lineto}}
\put(517,475){\special{em:lineto}}
\put(531,481){\special{em:lineto}}
\put(544,487){\special{em:lineto}}
\put(558,493){\special{em:lineto}}
\put(571,499){\special{em:lineto}}
\put(585,505){\special{em:lineto}}
\put(598,511){\special{em:lineto}}
\put(612,517){\special{em:lineto}}
\put(625,523){\special{em:lineto}}
\put(639,528){\special{em:lineto}}
\put(652,534){\special{em:lineto}}
\put(666,539){\special{em:lineto}}
\put(679,545){\special{em:lineto}}
\put(693,550){\special{em:lineto}}
\put(706,556){\special{em:lineto}}
\put(720,561){\special{em:lineto}}
\put(733,566){\special{em:lineto}}
\put(747,571){\special{em:lineto}}
\put(760,576){\special{em:lineto}}
\put(774,581){\special{em:lineto}}
\put(787,586){\special{em:lineto}}
\put(801,592){\special{em:lineto}}
\put(814,596){\special{em:lineto}}
\put(828,601){\special{em:lineto}}
\put(842,606){\special{em:lineto}}
\put(855,611){\special{em:lineto}}
\put(869,616){\special{em:lineto}}
\put(882,621){\special{em:lineto}}
\put(896,625){\special{em:lineto}}
\put(909,630){\special{em:lineto}}
\put(923,635){\special{em:lineto}}
\put(936,639){\special{em:lineto}}
\put(950,644){\special{em:lineto}}
\put(963,648){\special{em:lineto}}
\put(977,653){\special{em:lineto}}
\put(990,657){\special{em:lineto}}
\put(1004,662){\special{em:lineto}}
\put(1017,666){\special{em:lineto}}
\put(1031,670){\special{em:lineto}}
\put(1044,675){\special{em:lineto}}
\put(1058,679){\special{em:lineto}}
\put(1071,684){\special{em:lineto}}
\put(1085,688){\special{em:lineto}}
\put(1098,692){\special{em:lineto}}
\put(1112,696){\special{em:lineto}}
\put(1125,700){\special{em:lineto}}
\put(1139,705){\special{em:lineto}}
\put(1152,709){\special{em:lineto}}
\put(1166,713){\special{em:lineto}}
\put(1179,717){\special{em:lineto}}
\put(1193,721){\special{em:lineto}}
\put(1206,725){\special{em:lineto}}
\put(1220,729){\special{em:lineto}}
\put(1233,733){\special{em:lineto}}
\put(1247,737){\special{em:lineto}}
\put(1260,741){\special{em:lineto}}
\put(1274,745){\special{em:lineto}}
\put(1287,749){\special{em:lineto}}
\put(1301,753){\special{em:lineto}}
\put(1314,757){\special{em:lineto}}
\put(1328,761){\special{em:lineto}}
\put(1341,764){\special{em:lineto}}
\put(1355,768){\special{em:lineto}}
\put(1368,772){\special{em:lineto}}
\put(1382,776){\special{em:lineto}}
\put(1395,780){\special{em:lineto}}
\put(1409,783){\special{em:lineto}}
\put(1422,787){\special{em:lineto}}
\put(1436,791){\special{em:lineto}}
\sbox{\plotpoint}{\rule[-0.200pt]{0.400pt}{0.400pt}}%
\special{em:linewidth 0.4pt}%
\put(220,306){\special{em:moveto}}
\put(234,318){\special{em:lineto}}
\put(247,329){\special{em:lineto}}
\put(261,339){\special{em:lineto}}
\put(274,349){\special{em:lineto}}
\put(288,359){\special{em:lineto}}
\put(301,368){\special{em:lineto}}
\put(315,378){\special{em:lineto}}
\put(328,387){\special{em:lineto}}
\put(342,395){\special{em:lineto}}
\put(355,404){\special{em:lineto}}
\put(369,412){\special{em:lineto}}
\put(382,420){\special{em:lineto}}
\put(396,428){\special{em:lineto}}
\put(409,436){\special{em:lineto}}
\put(423,444){\special{em:lineto}}
\put(436,451){\special{em:lineto}}
\put(450,458){\special{em:lineto}}
\put(463,466){\special{em:lineto}}
\put(477,473){\special{em:lineto}}
\put(490,480){\special{em:lineto}}
\put(504,486){\special{em:lineto}}
\put(517,493){\special{em:lineto}}
\put(531,500){\special{em:lineto}}
\put(544,506){\special{em:lineto}}
\put(558,513){\special{em:lineto}}
\put(571,519){\special{em:lineto}}
\put(585,526){\special{em:lineto}}
\put(598,532){\special{em:lineto}}
\put(612,538){\special{em:lineto}}
\put(625,544){\special{em:lineto}}
\put(639,550){\special{em:lineto}}
\put(652,556){\special{em:lineto}}
\put(666,562){\special{em:lineto}}
\put(679,567){\special{em:lineto}}
\put(693,573){\special{em:lineto}}
\put(706,579){\special{em:lineto}}
\put(720,584){\special{em:lineto}}
\put(733,590){\special{em:lineto}}
\put(747,595){\special{em:lineto}}
\put(760,601){\special{em:lineto}}
\put(774,606){\special{em:lineto}}
\put(787,611){\special{em:lineto}}
\put(801,617){\special{em:lineto}}
\put(814,622){\special{em:lineto}}
\put(828,627){\special{em:lineto}}
\put(842,632){\special{em:lineto}}
\put(855,637){\special{em:lineto}}
\put(869,642){\special{em:lineto}}
\put(882,647){\special{em:lineto}}
\put(896,652){\special{em:lineto}}
\put(909,657){\special{em:lineto}}
\put(923,662){\special{em:lineto}}
\put(936,667){\special{em:lineto}}
\put(950,672){\special{em:lineto}}
\put(963,677){\special{em:lineto}}
\put(977,681){\special{em:lineto}}
\put(990,686){\special{em:lineto}}
\put(1004,691){\special{em:lineto}}
\put(1017,695){\special{em:lineto}}
\put(1031,700){\special{em:lineto}}
\put(1044,704){\special{em:lineto}}
\put(1058,709){\special{em:lineto}}
\put(1071,714){\special{em:lineto}}
\put(1085,718){\special{em:lineto}}
\put(1098,722){\special{em:lineto}}
\put(1112,727){\special{em:lineto}}
\put(1125,731){\special{em:lineto}}
\put(1139,736){\special{em:lineto}}
\put(1152,740){\special{em:lineto}}
\put(1166,744){\special{em:lineto}}
\put(1179,748){\special{em:lineto}}
\put(1193,753){\special{em:lineto}}
\put(1206,757){\special{em:lineto}}
\put(1220,761){\special{em:lineto}}
\put(1233,765){\special{em:lineto}}
\put(1247,769){\special{em:lineto}}
\put(1260,773){\special{em:lineto}}
\put(1274,778){\special{em:lineto}}
\put(1287,782){\special{em:lineto}}
\put(1301,786){\special{em:lineto}}
\put(1314,790){\special{em:lineto}}
\put(1328,794){\special{em:lineto}}
\put(1341,798){\special{em:lineto}}
\put(1355,802){\special{em:lineto}}
\put(1368,805){\special{em:lineto}}
\put(1382,809){\special{em:lineto}}
\put(1395,813){\special{em:lineto}}
\put(1409,817){\special{em:lineto}}
\put(1422,821){\special{em:lineto}}
\put(1436,825){\special{em:lineto}}
\end{picture}

\end{center}

\caption{
$T_{c}$, $T_{\rm BKT}^q$ and $T_{\rho} = T_{c}^{MF}$ versus noninteracting
fermion density.  The regions of the normal phase (NP), pseudogap phase (PP),
BKT phase and long-range order (LRO) phase are indicated.
We assumed that $m_{z}/m_{\perp} = 100$ and
$(m_{z}d^{2} |\varepsilon_{b}|)^{-1}=1$.}

\end{figure}

\end{document}